\documentclass{elsart}

\date{}

\usepackage[dvips]{graphicx}
\usepackage{amssymb}
\usepackage{url}

\begin{document}

\begin{frontmatter}



\title{Possible relation between galactic flat rotational curves and  the Pioneers' anomalous
acceleration}  \thanks{The published version can be found at {\em
New Astronomy}, \url{www.elsevier.com/locate/newast}. }


\author{E. Minguzzi}

\address{Department of Applied Mathematics, Florence
 University,
Via S. Marta 3, 50139 Florence, Italy}
\ead{ettore.minguzzi@unifi.it}

\begin{abstract}
We consider a generic minimal modification of the Newtonian
potential, that is a modification that introduces only one
additional dimensional parameter. The modified potential depends on
a function whose behavior for large and small distances can be fixed
in order to obtain (i) galactic flat rotational curves and (ii) a
universal constant acceleration independent of the masses of the
interacting bodies (Pioneer anomaly).  Then using a dimensional
argument we show that the Tully-Fisher relation for the maximal
rotational velocity of spiral galaxies follows without any further
assumptions. This result suggests that the Pioneer anomalous
acceleration and the flat rotational curves of galaxies could have a
common origin in a modified gravitational theory. The relation of
these results with the Modified Newtonian Dynamics (MOND) is
discussed.
\end{abstract}

\end{frontmatter}
%
%
%
%
%

\section{Introduction}
In 1998 Anderson et al. \cite{Anderson:1998jd} reported an
unmodeled constant acceleration towards the Sun of about $ a_P
\simeq 8.5 \times 10^{-8} \textrm{cm}/\textrm{s}^2$ for
 the spacecrafts Pioneer 10 (launched 2 March 1972), Pioneer 11
(launched 4 December 1973), Galileo (launched 18 October 1989) and
Ulysses (launched 6 October 1990).

In a subsequent report \cite{Anderson:2001sg} they discussed in
detail many suggested explanations for the effect and gave the
value $a_P = (8.74 \pm 1.33) \times 10^{-8}
\textrm{cm}/\textrm{s}^2$  directed towards the Sun.

The data covered many years staring in 1980 when due to the large
distance ($\simeq 20$ AU) of Pioneer 10 from the Sun the solar
radiation pressure became sufficiently small to look for unmodeled
accelerations. The data was collected up to 1990 for Pioneer 11
($\simeq 30$ AU) and up to 1998 ($\simeq 70$ AU) for Pioneer 10.

The  spacecraft masses were quite different. Galileo had a mass
$m_G=1298 \ \textrm{kg}$ (the orbiter mass at launch minus the
propellent) while the Pioneers had a mass $m_P=223 \ \textrm{kg}$.
Their orbits were also quite different, nevertheless the same
constant centripetal acceleration was detected. This acceleration
(the Pioneer anomaly) does not appear in the planet ephemerides or
in the accurate range measurement of the Viking mission
\cite{Anderson:2001sg}. It has been later confirmed that no such
anomalous centripetal acceleration affects the motion of the major
planets of the solar system \cite{iorio06,tangen06}.  If confirmed
the effect would imply a violation of the equivalence principle  as
heavy bodies with the mass of a planet would fall differently from
lighter bodies like the spacecrafts. However, the different masses
of the Galileo and Pioneer spacecrafts show that the anomalous
acceleration is independent of the mass of the free falling bodies
as long as they have a small mass, a fact that does not help in
clarifying why the planets are not subject to the anomalous
acceleration.
A systematic error would clearly solve this problem  but so far none
has been found. In has been suggested \cite{page06} that within some
years the observation of the motion of minor trans-Neptunian objects
could finally confirm or rule out, by  filling the observational
mass gap, the presence of  an anomalous acceleration.

In this work we shall consider the Pioneer anomalous acceleration as
real and, in order to avoid the difficulties for the different
behavior of light and heavy bodies, we shall consider `test
particles' of mass $m'$ in the field of an heavier body $m$, $m' \ll
m$. The concept of test particle we use is a relative concept
related to the ratio $m'/m$, thus, for instance, the Pioneer
spacecrafts are test particles in the field of the Sun ($\log
\frac{m}{m'}\simeq 28$), the stars at the outer edge of a galaxy are
test particles in the field of the galaxy ($\log \frac{m}{m'}\simeq
10 $) and the major planets of the Solar system are not test
particles ($\log \frac{m}{m'}\lesssim 8$). For a more precise
definition, that would clarify the origin of the assumed breaking of
the equivalence principle, one would need a more complete physical
theory than that provided in this work.



Soon afterwards Anderson's report it was noted by many authors
that there is a numerical coincidence between $a_P/c=(2.8 \pm 0.4)
\times 10^{-18} \textrm{s}^{-1}$ and the Hubble constant
\cite{Freedman:2000cf},
\[
H_0=(72\pm 8) \textrm{km}/(\textrm{s Mpc})=(2.3 \pm 0.3) \times
10^{-18} \textrm{s}^{-1}. \] This observation suggests that the
Pioneer acceleration could be a short range (at the size of the
solar system) universal acceleration, that is a constant
acceleration unrelated with the masses of the test particles and
the Sun. In this work we shall make this hypothesis
\begin{itemize}
\item[(i)] Test particles in the field of an heavy body of mass $m$ have at small distances an acceleration ${\bf
a}=a(r) {\bf \hat{e}}_{r}$  with $a(r)=-G\frac{m}{r^2}-a_P$, where
$a_P$ is a universal constant that does not depend on $m$, and $r$
is the distance between the test particle and the mass $m$.
\end{itemize}
Note that the physics underlying this assumption must be
complemented by defining the effect of the lighter body on the
heavier one.  The compatibility with the conservation of momentum
implies that the heavier body is attracted by the lighter one with
an acceleration $\vert a(r)\vert =G\frac{m'}{r^2}+\frac{m'}{m}
a_P$. Since $a_{p}$ is small and $m'/m$ is small the anomalous
acceleration on the heavier body would be undetectable.

Our second assumption is based on the observation that the
rotational curves of spiral galaxies are asymptotically flat. The
velocity of the stars and the hydrogen atoms far from the galactic
center does not fall as $v \sim \sqrt{Gm/r}$ as in Newtonian
gravity, instead it tends to a constant $v_{\infty}$. This fact is
usually explained with the presence of an invisible dark matter
halo around the galaxies which modifies the effective
gravitational potential. Let us  denote with ${\bf a}=a(r) {\bf
\hat{e}}_{r}$ the acceleration field generated by a mass $m$. Our
second assumption is
\begin{itemize}
\item[(ii)] $\lim_{r \to +\infty} r a(r)=-v^2_{\infty}=cnst. $
\end{itemize}
so that flat rotational curves follow.

\section{The modified potential}

The conditions (i) and (ii) are not satisfied by a Newtonian
potential, therefore we look for a minimal modification of the
Newtonian potential. By minimal modification we mean a
modification that introduces only one auxiliary dimensional
parameter. Without loss of generality the modified potential can
be written in the form
\begin{equation}
V(r)=-G\frac{m}{r} ( 1+f(\beta r)),
\end{equation}
where $\beta \in [L]^{-1}$ is the new dimensional parameter and
$f(x)$, $x=\beta r$, is a $C^2$ function $f: \mathbb{R} \to
\mathbb{R}$.

We are going to find some constraints on the function $f$ so as to
satisfy (i) and (ii). First, the Newtonian limit implies $f(0)=0$.
Note that since $V$ is defined only up to a constant, the function
$f(r)$ is defined only up to linear terms in $r$. Note also that
there is a rescaling  freedom in the definition of $f$ and $\beta$,
indeed let $\lambda \in \mathbb{R}-\{0\}$ and redefine
\begin{eqnarray}
\bar{f}(x)&=&f(\lambda x), \\
\bar{\beta}&=& \beta /\lambda,
\end{eqnarray}
then $\bar{f}(\bar{\beta}r)=f(\beta r)$. Under the assumption $\vert
f''(0)\vert \ne0$, we use this freedom to fix $\vert f''(0)\vert= 2$
and $\beta
>0$.

The acceleration field ${\bf a}=a(r) {\bf \hat{e}}_{r}=-{\bf
\nabla} V$ is given by (the derivatives are with respect to $x$)
\begin{equation} \label{pot}
a(r)=-G\frac{m}{r^2}(1+f)+G\frac{m \beta}{r} f'.
\end{equation}
 Let us consider the condition (i). Taylor expanding $f(x)$ and $f'(x)$ at $x=0$ we obtain
the acceleration field at small distances
\begin{equation} \label{tay}
a(r)=-\frac{G m}{r^2}+\frac{G m}{2} \beta^2 f''(0)+Gm\beta^2 O(\beta
r).
\end{equation}
The condition (i) is satisfied iff $f''(0)<0$, which due to our
normalization implies  $f''(0)=-2$,  and \begin{equation}\beta^{2}=
a_P/Gm\end{equation} where $a_P$ is a universal constant independent
of $m$. Consider the spacecraft Pioneer in  the solar system. The
last term on the right-hand side of Eq. (\ref{tay}), is much smaller
than the second one since $a_P/(Gm_{\odot}/d^2) < 10^{-3}$, where
$d<87AU$ is the Pioneer distance from the Sun and $m_{\odot}$ is the
mass of the Sun.

The condition (ii) implies
\begin{equation}
\lim_{r \to +\infty} \frac{r a(r)}{G m \beta}= \lim_{x \to
+\infty}(f'-f/x)=-\frac{v^2_{\infty}}{G m \beta} .
\end{equation}
Note that $x$ is a dimensionless parameter, it follows that as $r
\to \infty$, $f(\beta r) \to f_{\infty}(\beta r)$ a function that
solves the differential equation
\begin{equation} \label{ed}
f'_{\infty}-f_{\infty}/x=-K_1, \qquad K_1 \in \mathbb{R}^{+},
\end{equation}
and $v^2_{\infty}=K_1 G m \beta$. Using the relation between
$\beta$ and $a_P$ we obtain the Tully-Fisher relation
\begin{equation} \label{cds}
v^4_{\infty}=(K_1^2  a_P G) m,
\end{equation}
which expresses the proportionality between the mass (and hence
the luminosity) of the spiral galaxy and the fourth power of the
asymptotic rotational velocity.

We recall that,  more generally,  the Tully-Fisher relation states
$L \propto v_{\infty}^{p}$ where $L$ is the luminosity of the
galaxy. Observationally the wave-band dependent exponent $p$ stays
in the range $[2.5,5]$, and has the smallest scatter in the near
infrared for which $p$ is found to be close to 4 (see
\cite{mcgaugh00}).

The solution of Eq. (\ref{ed})  is
\begin{equation}
f_{\infty}=K_2 \beta r - K_1  \beta r \ln (\beta r),
\end{equation}
where $K_2$ is an integration constant. As observed previously we
may redefine the potential so that $K_2=0$. Alternatively the linear
freedom of $f$ can be fixed requiring $f'(0)=0$. We make the latter
choice. The constant $K_1$ is expected to be of the order of unity
since it is the finite limit of a dimensionless function
$f(x)/(-x\ln x)$ that comes from a yet unknown gravitational theory.
As a consequence the proportionality constant in the Tully-Fisher
relation is related to the Pioneer acceleration and therefore,
according to our model, should be of the order of $H_0 G c$. It has
long be recognized that the proportionality constant in the
Tully-Fisher relation has exactly the predicted magnitude
\cite{milgrom83b}.

\section{The relation with MOND}

It is interesting to explore how the modified potential dynamics is
related to the MOND theory \cite{milgrom83a,milgrom83b,saunders02}.
Let us introduce the Newtonian acceleration $g_N=Gm/r^2$, the MOND
characteristic acceleration $a_0=K_1^2 a_P$ and the function $z(y)$,
(with $y=1/(K_1 x)^2={g_N}/{a_0}$)
\begin{equation} \label{er}
z(y)=y[1+f(\frac{1}{K_1\sqrt{y}})]-\frac{\sqrt{y}}{K_1}\,
f'(\frac{1}{K_1\sqrt{y}}),
\end{equation}
then Eq. (\ref{pot}) can be rewritten
\begin{equation}
{a}/{a_0}=-z(g_N/a_0).
\end{equation}
For problems with spherical or cylindrical mass configurations the
theory reduces to MOND provided $z(y)$ has an inverse
$I(z)=z\mu(z)$, $I(z(y))= y$, with $\mu(z) \sim z$ for $z\ll 1$
and $\mu(z)\sim 1$ for $z \gg 1$.

Let us see whether these conditions are compatible with the modified
potential. The limit $x \to +\infty$, corresponds to $y \to 0$ and
the asymptotic behavior of $f(x)$ implies that $z \to 0$ as $z(y)
\sim \sqrt{y}$ and hence $\mu(z) \sim z$. The limit $x \to 0$
corresponds to $y \to +\infty$ and $f(0)=0$ implies that $z \to
+\infty$ as $z(y) \sim y$ and hence $\mu(z) \to 1$. Thus for
symmetric mass configurations we recover MOND.

Some comments are in order. The function $z(y)$ does not necessarily
need to be invertible for certain choices of $f(x)$, hence a
`modified inertia' formulation in terms of a function $\mu$ is not
guaranteed. The asymptotic behavior of $f(x)$ for $x \to 0,+\infty$
used above does not suffice to recover MOND from the minimally
modified potential. Indeed, we used also condition (i) that led to a
constraint for $\beta$ and then to the functional form (\ref{er})
for $z(y)$. In our modified potential formulation the Tully-Fisher
relation is derived from the unrelated assumption (i) while in MOND
it follows by construction from the condition $\mu(z) \sim z$ for
$z\ll 1$. MOND does not give necessarily  the Pioneer anomaly that
instead is accommodated since the beginning in our formulation. In
MOND the Pioneer anomaly would be included imposing the additional
constraint $f''(0)=-2$. It is not difficult to show that it
corresponds to
\begin{equation} \label{con}
\mu(z) \sim 1-\frac{1}{K_1^2 z},
\quad \textrm{as} \quad z \to +\infty.
\end{equation}
Since the differences between MOND and our derived dynamics are
only minimal we can regard these calculations as a proof that a
MOND theory subject to constraint (\ref{con}) follows from
assumptions (i) and (ii). Note that MOND does not satisfy the
equivalence principle a fact which is coherent with the
phenomenology related to the Pioneer anomaly from which we
started.

The observations give (through Eq. (\ref{cds})) $a_{0}=1.2 \times
10^{-8} \textrm{cm}/\textrm{s}^2$, and we find from the relation
$a_0=K_1^2 a_P $ that $K_1^{-2} \simeq 7$. We stress again that
since $K_1$ is of the order of unity the dimensional argument is
sound.

Some work has been done to restrict the function $\mu(z)$ of MOND
theory. In particular we may ask whether the constraint
(\ref{con}) is compatible with the observations. Sereno and Jetzer
\cite{sereno06} pointed out that a similar asymptotic behavior is
incompatible with the accurate data available on Mars orbit.
However, this is not a surprise, since in order to take the
Pioneer anomaly seriously  we assumed since the beginning that the
equivalence principle has to be violated. Clearly the chosen $\mu$
function, or equivalently the $f$ function, can not fit the
Newtonian orbits of the major planets since it has been chosen in
order to reproduce the Pioneer anomaly which, as we said in the
introduction, does not affect the orbit of those planets. We had
to make a choice: whether to fit the $f$ function to reproduce a
typical Newtonian behavior or to fit it with the data on the
Pioneer anomaly. We chose the latter possibility so as to be in a
small test particle limit. The choice to consider the problem in
some limit makes sense taking into account that the theory was
already expected to violate the equivalence principle.

Apart from the constraint (\ref{con}) the function $\mu$ of our
theory is a typical $\mu$ function of MOND theory.  Famaey and
Binney \cite{famaey05,zhao06} showed that the simple choice
$\tilde{\mu}(z)=z/(1+z)$ is particularly successful in fitting the
Milky Way and the galaxy NGC3198, in particular, it proved
superior than the traditional choice $\check\mu=z/\sqrt{1+z^2}$.
It is interesting to note that the function $\tilde\mu$ has the
asymptotic behavior $\tilde\mu \sim 1-\frac{1}{z}$, and hence,
although it is not compatible with our theory (as the expansion is
slightly different from the one required) it implies the presence
of an anomalous centripetal acceleration of value $a_0$.
Nevertheless, a fit of the Milky Way led Famaey and Binney
\cite{famaey05} to conclude that the function $\tilde{\mu}$ gives
good results only up to values $z\lesssim 5$, while for the
Pioneers we are in the range $z \sim 10^3$. At that range Famaey
and Binney argue that a transition should have already taken place
to the function $\check\mu$, which unfortunately does not imply a
universal centripetal acceleration. Although, these results are
not conclusive, as they are based on the study of only two
galaxies, they seem to imply that the Pioneer anomalous
acceleration does not show up in the dynamics of galaxies.

We end the section by noticing  that since the Pioneer anomaly is
mainly a post-Newtonian effect it was natural to start our study
from Galilei invariant assumptions (i) and (ii). Nevertheless, the
closeness of the final theory to MOND and the possibility of
generalizing MOND in a relativistic way \cite{bekenstein04}
clarifies that those assumptions were not incompatible with
relativistic physics as long as they are taken in the suitable
slow-speed weak-gravity limit.


\section{Conclusions}
The galactic flat rotational curves and the Pioneer anomaly are
among the few  phenomena that could suggest a departure from the
Newtonian gravitational potential. In this work we  assumed a
common origin in a modified but yet unknown (effective)
gravitational theory. We considered the case of two masses, $m'
\ll m$, and introduced a minimally modified potential, that is a
potential that involves only one auxiliary dimensional parameter.
We showed that such potential can be suitably tuned to produce the
said phenomena. We found that a minimally modified potential
$V(r)$ that meets conditions (i) and (ii) has the form
\begin{equation} \label{kju}
V(r)=-G\frac{m}{r} [ 1+f(\sqrt{\frac{a_P}{Gm}}\, \,r)],
\end{equation}
where function $f(x)$ satisfies $f(0)=0$, $f'(0)=0$, $f''(0)=-2$,
and has the asymptotic behavior $f \sim -K_1  x \ln x$. Functions
of this kind exist, consider for instance the simple choice,
$f(x)=-K_1 x \ln (1+x/K_1)$. Moreover, the asymptotic rotational
velocity is related to $K_1$ by $K_1=\frac{v_{\infty}^2}{\sqrt{a_P
G m}}$, and hence the Tully-Fisher relation holds. The emergence
of the Tully-Fisher relation can be considered as a suggestion
that the starting assumption, i.e. that  the flat rotational
curves and the Pioneer anomaly have the same gravitational origin,
could indeed be correct. Nevertheless, we showed that under the
said assumptions we recover a MOND like theory subject to the
constraint (\ref{cds}). Most of the successful predictions of MOND
theory do not depend on the particular form of the $\mu(z)$
function (i.e. on $f(x)$), however, in the last years some work
has been done to constraint the function $\mu$ by using galactic
rotation curves. Using this data and results by other authors we
concluded that it is unlikely that the Pioneer anomalous
acceleration shows up at the galactic scales. In other words
although the assumption of the Pioneer anomaly naturally leads to
the the successful MOND theory,  the latest information available
on galaxies seems to exclude the presence of an anomalous
centripetal acceleration of the order of $a_P$. This complex
situation seems to require more investigation, the possibility
 of a relation between galactic dynamics and the Pioneer anomaly being still open.

\section*{Acknowledgments} It is a pleasure to thank  V.J. Bol\'os
and J.-F. Pascual-S\'anchez for sharing many useful discussions on
the Pioneer anomaly. Work supported by INFN, Grant No. 9503/02.


\end{document}